# Ninth Planet or Wandering Star ?

by
Gilles Couture
Département des Sciences de la Terre et de l'Atmosphère
Université du Québec à Montréal
CP 8888, Succ. Centre-Ville
Montréal, QC, Canada, H3C3P8
(couture.gilles@uqam.ca)

**Abstract**

We study the gravitational effects of two celestrial bodies on a typical object of the Kuyper Belt. The first body is a kuyperian object itself with fairly large eccentricity and perihelion but with a large mass, about 16 times the mass of the Earth. The second body is a star whose mass is 30% – 50% of the mass of the sun that passes by our solar system at a speed between 25 km/sec and 100 km/sec and at a distance of closest approach between 0.05 and 0.5 light year. As a measure of the perturbations caused by these bodies on the light kuyperian object, we analyse its eccentricity. We find that the effects due to the passage of the wandering star are permanent in the sense that the eccentricity of the kuyperian object remains anomalous long after the passage of the star. The same is true of the heavy kuyperian opbject: it can perturb greatly the orbit of the lighter kuyperian objet, which leads to a permanent, anomalous eccentricity.

## Introduction

The discovery of Sedna [1] was a very exciting event in modern astronomy. Several objects of the Kuyper Belt had been discovered since the early 1990's but Sedna was different. Not only was it farther out than any of the previous objects, its orbit was also more eccentric than anything observed before, apart from comets. Its aphelion was by far the longest observed in the solar system. Since this first observation, a dozen or so of these objects have been discovered. Sedna however remained in a class by itself due to its unique orbital parameters. The discovery of 2012 $VP_{113}$ [2] was special because this object had orbital parameters similar to those of Sedna. To this day, these two objects remain in class by themselves when considering their orbital parameters. [3]

Such peculiar objects raise several questions regarding their genesis [4]: did they form at such large distances from the sun or were they formed much closer to the sun and then moved to their current orbits ? What mechanism could take such an object from a relative proximity to the sun and transport it to such far distances ? [5,6] If this mechanism has been used for Sedna and its siblings, (also called Sednitos) could we expect many more similar objects ?

When studying these objects, it has been noticed that they tend to cluster in a relatively small section of the outer solar system. This is intriguing and has raised the possibility of an object that, while belonging to the same class of objects, would be substantially more massive.[7] It has been estimated that in order to be able to shepherd these objects in that part of the solar system, this massive object would have a mass of about 16 Earth masses, and an orbit of comparable eccentricity (about 0.6) with an aphelion of about 120 AU. This would be the Ninth Planet (NP) of our solar system and current planetary models allow us to estimate its structure [8].

This is a very interesting possibility that has attracted a lot of attention over the past 2 years. Several groups are now searching for this massive Kuyperian object. There are some hints of its wherabouts. Shortly after NP was suggested, a fairly large part of its putative orbit was ruled out through a careful analysis of Cassini data; although very far from us, NP would have a small effect on Saturn [9]. A mean motion resonance mechanism between NP and other Sednitos was suggested as the source of the clustering of the Sednitos' orbits in [7] and more detailed studies followed [10,11]. Furthermore, the effect of NP on the mass distribution within the Kuyper belt has been calculated [12] and while the distributions with and without NP are qualitatively different, it could be very difficult to distinguish them observationally.

While exciting, the hypothesis of a ninth planet is still debated and will remain so until NP is discovered. Recently, it has been argued that the orbits of some Sednitos have been wrongly estimated, thereby weakening the reason to invoke NP at all.[13]; although recent results tend to agree with the presence of a ninth planet [14] There is also the possibility that all Sednitos were captured by the sun during a close encounter with a star and its own planets.[15]

About four years ago [16], a faint binary system consisting of a small red dwarf and a smaller brown dwarf was discovered and named WISE J072003.20-084651.2. It is also commonly refered to as Scholz's star, from the name of its discoverer at Leibniz-Institut für Astrophysik in Potsdam. This object was peculiar as it did not seem to follow the motion of the nearby stars; it is a wandering star. Furthermore, it did not move at all in the the sky, as if it was coming from the solar system. About two years ago, carefull measurements [17] showed a very small angular motion that indicates that this star was at some point in the past very close to our solar system. The origin of this particular WS is still not well understood, but we know of some mechanisms that can produce wandering stars. A double star system that gets too close to a cluster of stars within a galaxy could be broken up by tidal forces and produce two stars on wandering paths. At the center of a galaxy, if a binary system gets too close to a black hole, it is possible for one star to fall into the black hole while the other star is ejected on a new trajectory. One could even think of one star being simply extracted from its galaxy during a close encounter between two galaxies and, not being able to catch up with the visiting galaxy, ends up as a wandering star between galaxies. Scholtz's star is now about 20 LY from us and from its speed of about 100 kms/sec, it is estimated that it came well within 1 LY (about 0.75 LY) to us about 70000 years ago. The masses of the two companions are estimated at 8% and 6% of a solar mass.

In this work, we want to compare the effects of a WS to those of a NP on a typical kuyperian object and see whether we could distinguish between these two encounters by analysing the motion of the perturbed object.

**Calculations**

In what follows, we will compare the unperturbed orbit of a typical kuyperian object to its orbit perturbed by a heavy object. If one had access to the unperturbed orbit, it would be very simple to measure such discrepancies. However, when we look up in the sky, we do not have the unperturbed orbit, we have the real orbit that might be perturbed or not. As a measure of the perturbations caused by the heavier objects in the problem it seemed natural to analyse the eccentricity of the path of the reference object. Without perturbation, this should remain constant throughout the path. If the orbit has been perturbed by an object other than the sun, the eccentricity should not be constant all along the orbit. As we will see, the eccentricity of the orbit remains irregular for a long time once it is perturbed either by another heavier kuyperian object or a wandering star. As the eccentricity is difficult to measure or calculate accurately at perihelion and aphelion, we will drop a small portion of the orbit at $\pm 1^o$ of these positions.

The typical kuyperian object that we will use in our work is a Sedna-like object (SO): very small mass compared to the Earth, perihelion of $10^{13}$ m and eccentricity of 0.85 which gives an aphelion of $1.233 \times 10^{14} m$ and a period of about 9383 years ($2.96118 \times 10^{11}$ sec). We will assume that the ellipse is initially on the horizontal axis (x-axis) with the CM of the system (Sun-SO) at the origin and the perihelion on the positive x-axis. The rotation will be counterclockwise starting at $\theta = 0$ at perihelion. Because of its large eccentricity, it will use about 20% of its period to travel from $(0, y_{min})$ to $(0, y_{max})$ and about 80% of its time to go from $(0, y_{max})$ to $(0, y_{min})$. This kuyperian object will always start its motion at perihelion.

We will assume that the NP rotates counterclockwise also. The mass that we will use for this object is $10^{26}kg$, about 16 times the mass of the earth. Then we will consider first a perihelion of $1.2 \times 10^{13}$m and an eccentricity of 0.75, which leads to an aphelion of $8.4 \times 10^{13}$m and a period of about 5733 years. We will also consider n perihelion of $3 \times 10^{13}$ with and excentricity of 0.70, which leads to an aphelion of $1.7 \times 10^{14}$m and a period of about 17250 years . In both cases we will vary the orientation of the orbit and the starting position of the object. When the orbit of this object is not aligned with that of SO, we will measure the angle ($\phi$) between the semi-major axes clockwise from the horizontal. For example, if the orbit is vertical, it will be at an angle of $\phi = \pi/2$ and its perihelion will be on the negative y-axis. When this object starts its motion at its perihelion, we will say that $\theta_{NP} = 0$ and if it starts at its apelion, we will say that it sarts at $\theta_{NP} = \pi/2$. Once the initial conditions are set, we use analytical equations for the first few time steps and then we simply let the system evolve according to the force equations.

The parameters that we will use for the WS are a mass of $(0.3-1.0) \times 10^{30}$ kg (about 15% – 50% of the mass of the sun) and an incoming speed of 25 km/sec and up to 100 kms/sec initially on a linear trajectory. We will vary its initial positions along x and y axes in order to see the effects of these parameters on the eccentricity of SO. This allows us to vary the timing of its encounter with SO on its orbit. This object will feel the pull of the sun and will end up closer to it than its initial transversal position.

The reader should keep in mind that $1LY = 0.947 \times 10^{16}m \sim 10^{16}m$. Therefore, $0.3LY \sim 0.3 \times 10^{16}m$.

As we are dealing with a problem in three dimensions, the combinations of encounter parameters are limitless. We will limit our study to a two-dimensional problem. This two dimensional problem will represent properly the full problem and give a very good idea of the perturbations that would result from the full three dimensional encounters.

The first step of our calculation is to numerically solve the equations of motion of our SO arising from the different forces in the problem. In order to do so, one needs the position and velocity at a given instant in order to calculate the acceleration, new velocity and the distance traveled during a given time interval. We simply use the analytical solution to calculate these first few time steps and then we let the system evolve according to the force equations. One then needs a time scale, a time step. Given our numerical capabilities and the stability of the orbits required, we found that a time step of about 317 seconds was appropriate: this gave stable accurate numerical solutions and required a reasonable amount of time. In order to verify the accuracy of our numerical solutions, we compared the results to the exact, analytical solution. We let the object evolve on several orbits and verify the stability of the numerical solutions and the discrepancy between the numerical and analytical solutions. Regarding the stability of our numerical orbits, we find that there is a wandering of about 1 part per million over several orbits: the positions at apelion of the SO differ by 1 part per million over several orbits. When we compare our numerical orbits to the analytical solution, we find that they differ also by 1 part per million over several orbits.

The second step is to add the forces coming from a third object and see how the presence of this third object will affect the orbit of the SO. Since the three-body problem does not have a general solution, this part is completely numerical.

The **first third object** that we consider is NP. Since its mass is much smaller than the mass of the sun, its effect on the sun is again calculable, but totally negligible from a numerical point of view. The first scenario considered here is a perihelion of $1.2 \times 10^{13}$m with an eccentricity of 0.75, which leads to an aphelion of $8.4 \times 10^{13}$ m. As this perihelion is close to that of the SO, the effect of this object will be greatest when it is closest to SO: when it starts at $\theta = 0$ its orbit has the same orientation to that of the SO. In these conditions, the effects are huge and greatly perturb the eccentricity of SO, as we can see on figure 1. We also note a nice resonance phenomenon that brings back SO to a more stable eccentricity, still not a constant eccentricity before returning to a highly variable eccentricity. Note that in this figure, as with all figures, the time span between two points is about 130 years. The effects are much less when NP starts at aphelion ($\theta = \pi$) but tend to increase with time as NP and SO happen to pass close by eachother. We can see also that these effects are permanent: the eccentricity will never be constant after such a close encounter. Therefore, if such an object were to have come close to SO in the distant past, the eccentricity would still be affected and highly irregular.

On figure 2, we show the effect of NP when its orbit is not aligned with that of SO. In the first case, it is at an angle of 225 with respect to SO and NP starts at its perihelion. In the second case, its orbit is aligned with that of SO but it starts at its aphelion. In both cases, we see that the effect is small at the beginning (since the two objects are far from each other) but grows with time. Clearly, at some point in time they will get close to each other and the perturbations on SO will be large.

On figures 3a and 3b, we see the effect of a NP that is farther away from SO: its perihelion is either $(3, 6, 10) \times 10^{13}$ m. All these have an eccentriticy of 0.70, which leads to aphelions of $(17, 34, 56.7) \times 10^{13}$ m, respectively. All these perturbations are substantially smaller than previously. One notes however that even at $3 \times 10^{13}$, the perturbations are noticeable both for the aligned and misaligned orbits. One can presume that for the misaligned orbits, there will come a point when the two orbits will cross and the two objects will be close to each other. This will result in fairly large perturbations on SO. The last two cases lead to rather small perturbations because the aphelions are much larger than SO's and they will never cross since they are aligned. Again, if these orbits are misaligned with SO, the paths will cross and the perturbations will eventually get large.

We conclude here that when a heavy kuyperian object has a fairly close encounter with a much lighter kuyperian object, it will affect its path forever and this will lead to a variable eccentricity along the orbit. In general, this effect will be fairly large. In order to have a small effect, NP must stay far away from SO at all time. This requires a substantially larger perihelion, a smaller eccentricity and an orbit that does not cross that of SO. Otherwise, if the orbits cross, the two objects will eventually get close enough for the heavier one to have a large and everlasting effect on the eccentricity of the lighter object.

The **second third object** that we consider is the WS. Current data on this object indicate that it was closest to our solar system about 70,000 years ago and was then well within 1 light-year from the sun. Since then, it has been moving at a speed of about 100 km/sec. Its mass is about 15% - 30% of the solar mass. Since this third object has a mass comparable to that of the sun, it will pull the sun out of its regular orbit. However, this object passed by our solar system long before humans started making measurements of the stars and planets. Therefore, this perturbed motion of the solar system is what we have been measuring for centuries; it is our motion around the galaxy. We will take this into consideration simply by calculating the motion of the SO not with respect to our absolute frame but with respect to the sun. We assume that this object has no effect on the motion of the earth around the sun. Therefore, by taking the relative motion of the SO with respect to the sun, we are in fact taking the relative motion of the SO with respect to the earth. Again, we will use the excentricity of the orbit as a measure of the perturbation caused by the passage of this WS. We expect this effect to produce an excentricity that will vary along the orbit.

In order to allow for this wandering star to get more or less close to the sun and meet SO at different positions on its orbit, we will vary its initial starting position $(x_0, y_0)$ and vary its initial velocity $(v_0^x, v_0^y)$. Its initial transverse position to the sun will vary from $(0.05 - 0.5) \times 10^{16} m$ and its initial velocity will vary from 25km/sec to 100 km/sec. For comparison, SO's aphelion is about $1.233 \times 10^{14}$m; the WS stays always fairly far from the solar system.

An interesting result it that when WS is moving perpendicular to the ellipse of SO, its effect on SO's eccentricity is largest typically when it reaches the major-axis of SO's orbit when SO has completed only one quater of its orbit (which takes about 10% of its period) or has completed about 3/4 of its orbit (which takes about 90% of its period) When WS is moving parallel to SO's orbit, its effect on SO's eccentricity will be largest roughly when it crosses the minor-axis of SO when SO has completed about 1/4 of its orbit. Also interesting is that the effect of WS is minimal when it crosses the major-axis of SO's ellipse when SO is at perihelion or aphelion; the whole process is then perfectly symmetric in time.

On figure 4-7, we present the excentricity of the orbit of the SO over a long period of time. In this scenario, the WS reached the aphelion of the orbit $3.28691 \times 10^{12}$ seconds after it started moving towards the solar system at a speed of 100 km/sec. It started its journey at a distance of $32.8691 \times 10^{16} m$ (about 32 LY) on the positive or negative y axis. At this speed and initial position, it reached the aphelion of the SO orbit when SO had completed about 11.1 cycles. These figures show that the WS has a long lasting effect on the orbit of the SO. In fact, the orbit of the SO will never be an ellipse anymore and forever will it have a variable excentricity along its orbit.

On figure 4, we see the effect of varying the initial transverse distance of WS and its mass on SO's eccentricity. Clearly, the larger the mass and the close it gets to SO, the stronger its effect is.

On figure 5, we see the effect of varying the distance while keeping the other parameters constant. Clearly, the effect becomes very small when the distance exceeds 0.3 LY.

On figure 6, we see the effect of the different times of arrival of WS. We see that the effect is approximately the same whether WS crosses x-axis when SO is at 11.1 orbits or 10.9 orbits. We also see the effects when WS runs parallel to the ellipse instead of perpendicular to it. When it reaches the y-axis with SO, the effect is larger and about the same whether they meet head-on on WS is running after SO.

On figure 7, we use a very large mass of $10^{30}$ kg for WS and see its effects can be substantial. We see that the amplitude of the perturbations varies a lot with the transverse distance of departure but not very much with the axis of motion: the red and magenta curves are very similar in amplitude and are almost mirror of each other.

On figure 8, we show the effect of varying the speed of the WS and its initial position transverse to its motion. In all cases, WS crossed the y axis when SO had travelled 11.1 cycles. Clearly, we see that a slower star has a stronger effect on SO, as one would expect.

Therefore, in order to have a strong effect on SO, WS must be as heavy as possible, pass as close to SO as possible and as slowly as possible.

An important fact to remember here, again, is that these perturbations are forever. The do not die away and are a signature of the passage of WS. This means that the passage of a wandering star affects a kuyperian object in a way that is similar to that produced by another, more massive kuyperian object such as NP.

**Conclusions**

We have studied the perturbations on a typical kuyperian object caused by a heavy kuyperian object or a wandering star. We have seen that both encounters have ever lasting effects on the light object and that those from a heavy kuyperain object can be very large if it crosses the orbit of the light object. One signature of such encounters is an anomalous eccentricity, an eccentricity that is not constant throughout the path. In general, the anomaly in the eccentricity is largest at perihelion and aphelion, then it decreases rather quickly and reaches a plateau for a large section of the orbit. Unless one can justify why kuyperian objects would never get close to one another, we have to conclude that small perturbations in the eccentricity of SO cannot come from a heavy kuyperian object of 16 earth masses; indeed, if these two objects come within $few \times 10^{13} m$, the eccentricity of the lighter object will vary by a few percent along its path. Such an object has to remain fairly far away from SO otherwise, the perturbations will be very large.

On the other hand, we have seen that a wandering star will have a relatively small effect in general on the eccentricity of a typical kuyperian object such as SO. We have to conclude that WISE J072003.20-084651.2 had a very small effect on kuyperian objects similar to Sedna, because of its rather small mass and fairly large distance of closest approach. If the object has a mass of about 0.3 solar mass or larger and gets closer than about 0.25 LY from the sun, then, depending on its relative position with respect to the SO, its effects can be more substantial. Clearly, such an object would be a bright star, which we would likely have observed already. However, one could think of a combinaison of several small objects such as brown dwarfs, red dwarfs or even white dwarfs and black holes whose total mass could be as large as 1 solar mass and would be difficult to detect because of their very small combined luminosity. Therefore, if the perturbations observed on SO are small, they are as likely to come from a heavy kuyperian object that orbits very far from the sun in order to remain always at large distances from SO as from wandering stars close to each other that come close to our solar system. Scholtz's star has been discovered just recently. Are there other similar objects that could have affected the orbits of other Kuyperian objects ?

Our calculations indicate that if a fairly massive kuyperian object has had an effect strong enough on several light kuyperian objects to shepherd them in a given section of our extended solar system, it is likely that these effects have been large enough to produce orbits with anomalous eccentricities. The anomalies should be of the order of a percent or more over the entire orbit and one could be able to meaure them over a period of 30 years or so. If the orbit of the heavy kuyperian object crosses the orbits of the lighter ones, such large perturbations are unavoidable.

As binary systems are fairly common [18,19,20], there is also the possibility that our own sun had a sister star in its early life that would have left us later on due to nearby stars. Clearly, such as sister star would have had a profound impact on the kuyperian objects that stayed with our sun and likely, these effects would remain today. It might be possible to use the orbits of our own kuyperian objects to know the past of our solar system since our results indicate that such ephemeral encounters do have ever-lasting effects on our kuyperian objects and the astronomical perturbations that they cause, like diamonds are forever.

**Acknowledgements**
It is a pleasure to thank my colleague Chérif Hamzaoui for interesting, enjoyable and stimulating discussions and Manuel Toharia for his visits and discussions. I also want to thank Mr. Anthony De La Llave for collaboration in the early stages of this work.

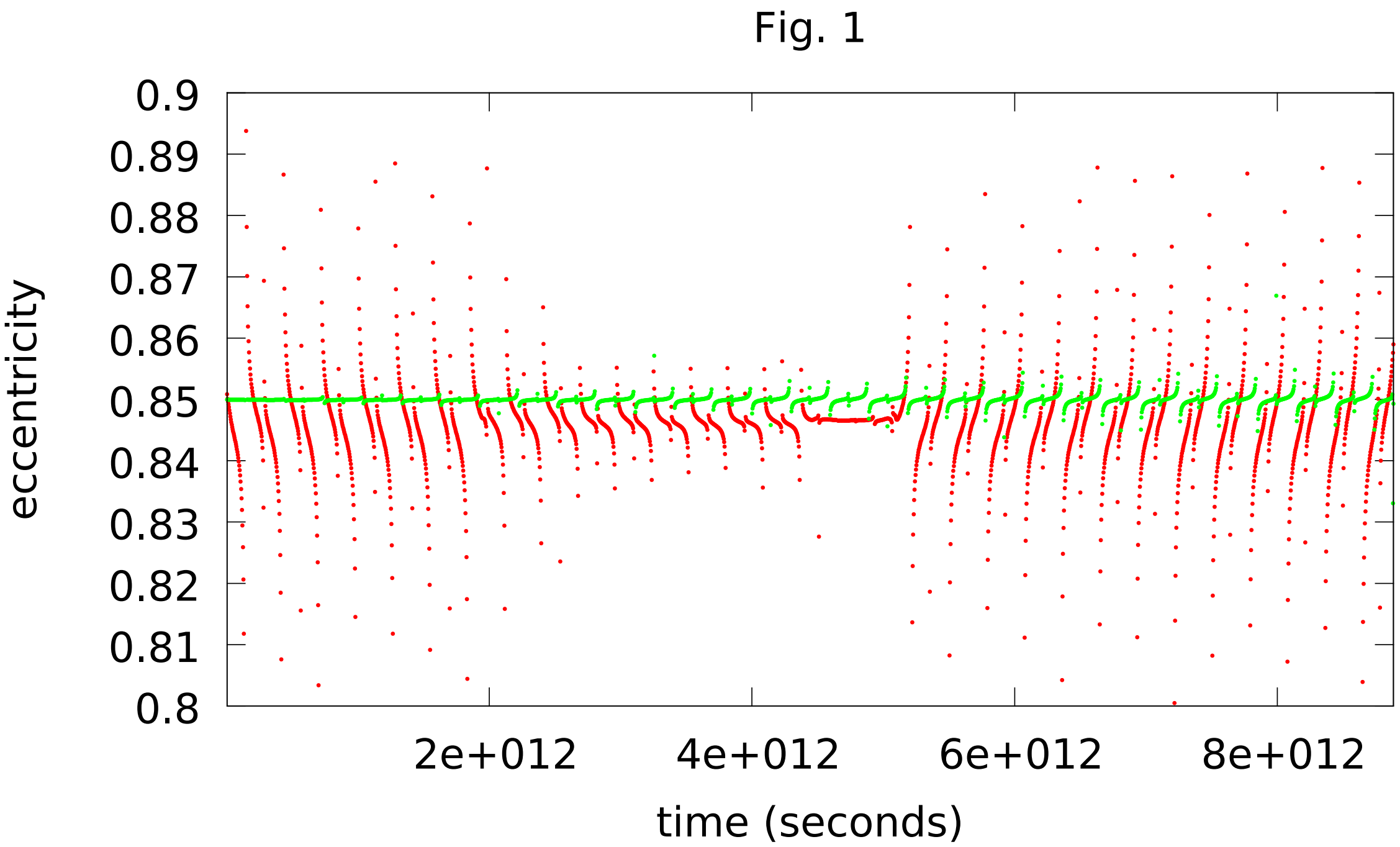

Fig. 1 NP perihelion is $1.2 \times 10^{13} m$ and $ecc = 0.75$. $\theta = 0$ and $\phi = 0$ in red and $\theta = \pi$ and $\phi = 0$ in green.

Fig. 2

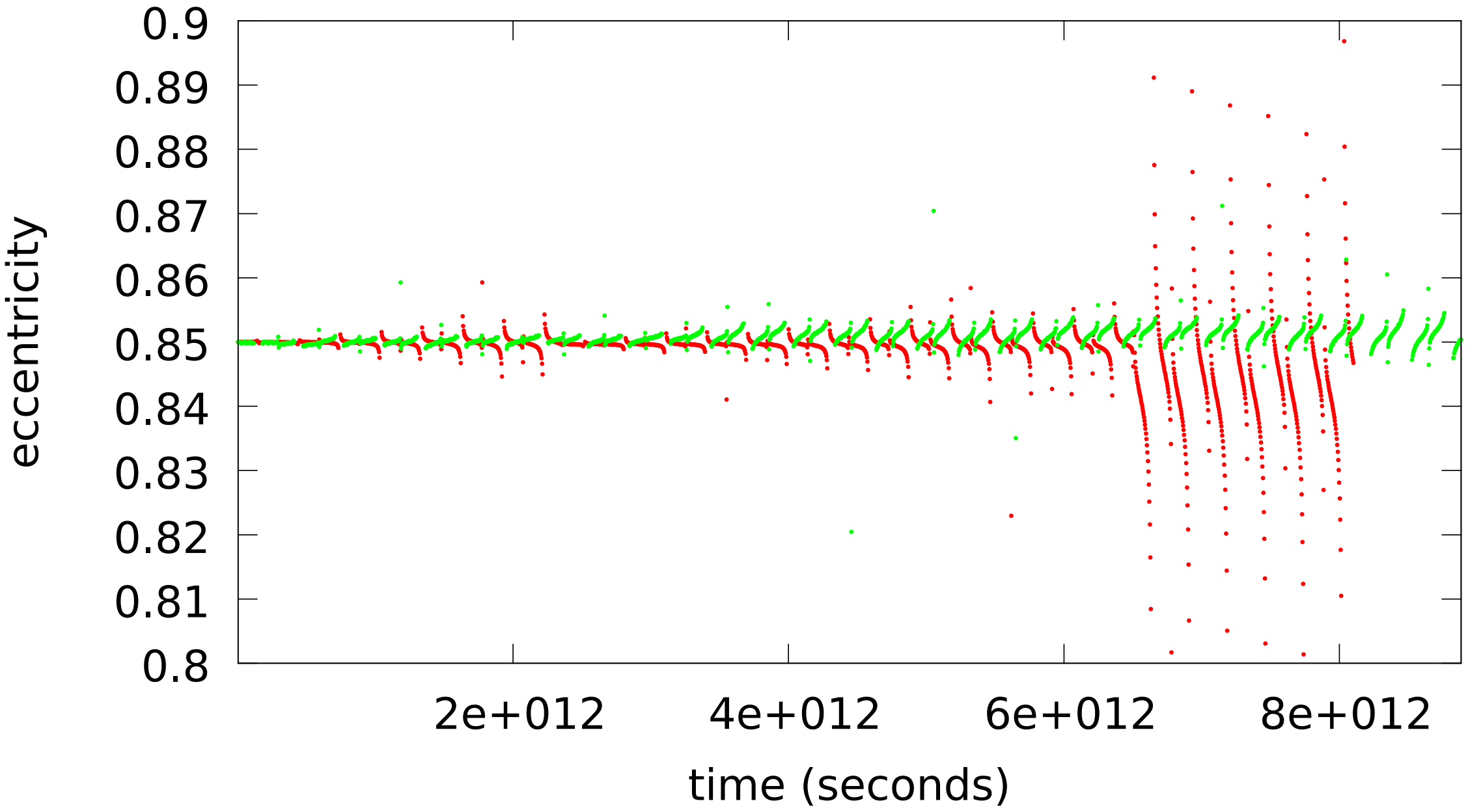

Fig. 2 NP perihelion is $1.2 \times 10^{13} m$ and $ecc = 0.75$. $\theta = 0$ and $\phi = 5\pi/4$ in red and $\theta = pi$ and $\phi = 3\pi/2$ in green.

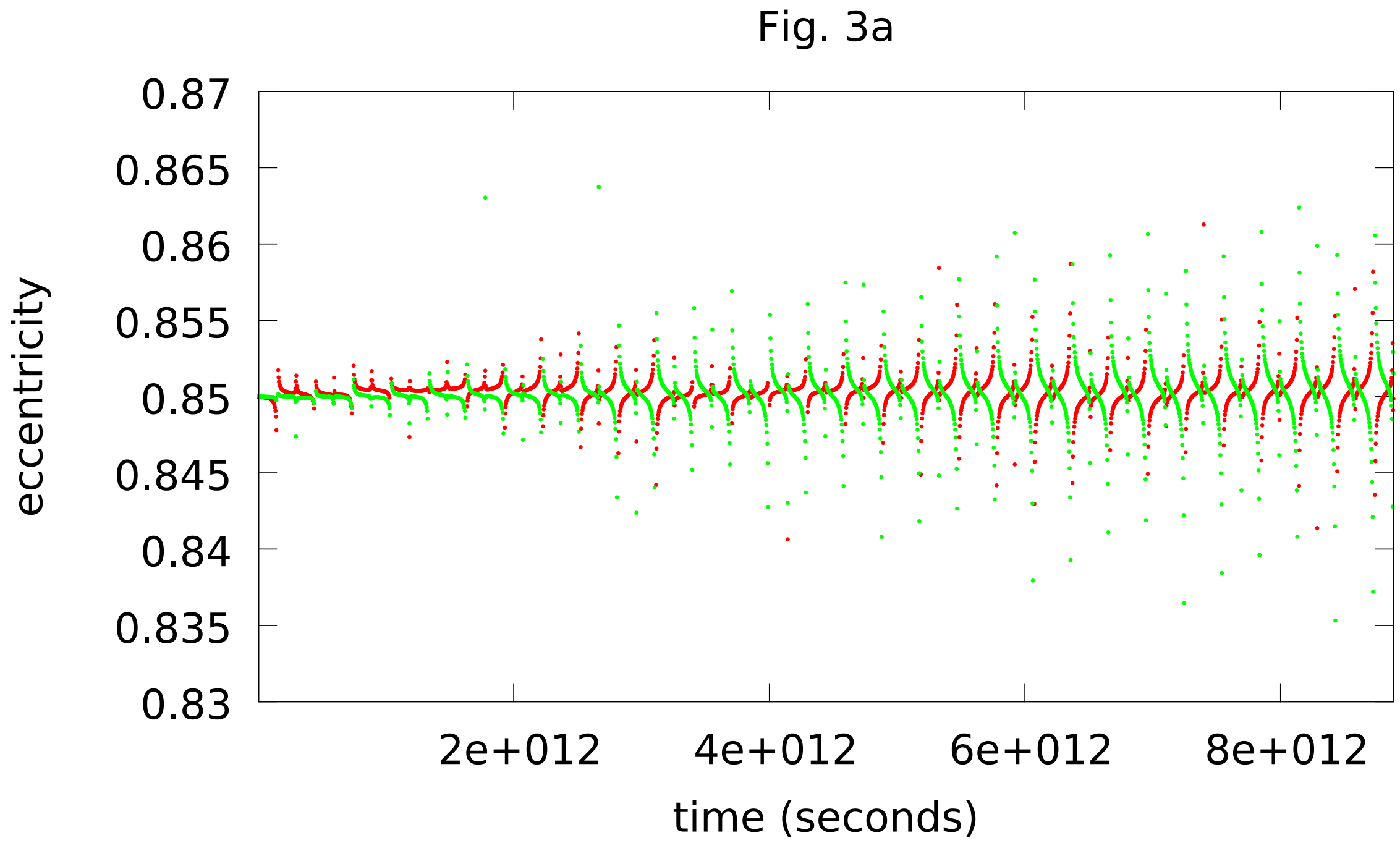

Fig. 3a $ecc = 0.70$ in this figure. NP perihelion is $3 \times 10^{13} m$ and $((\theta = 0 = \phi) in\ red\ and\ (\theta = 3\pi/4,\ \phi = 0)\ in\ green)$.

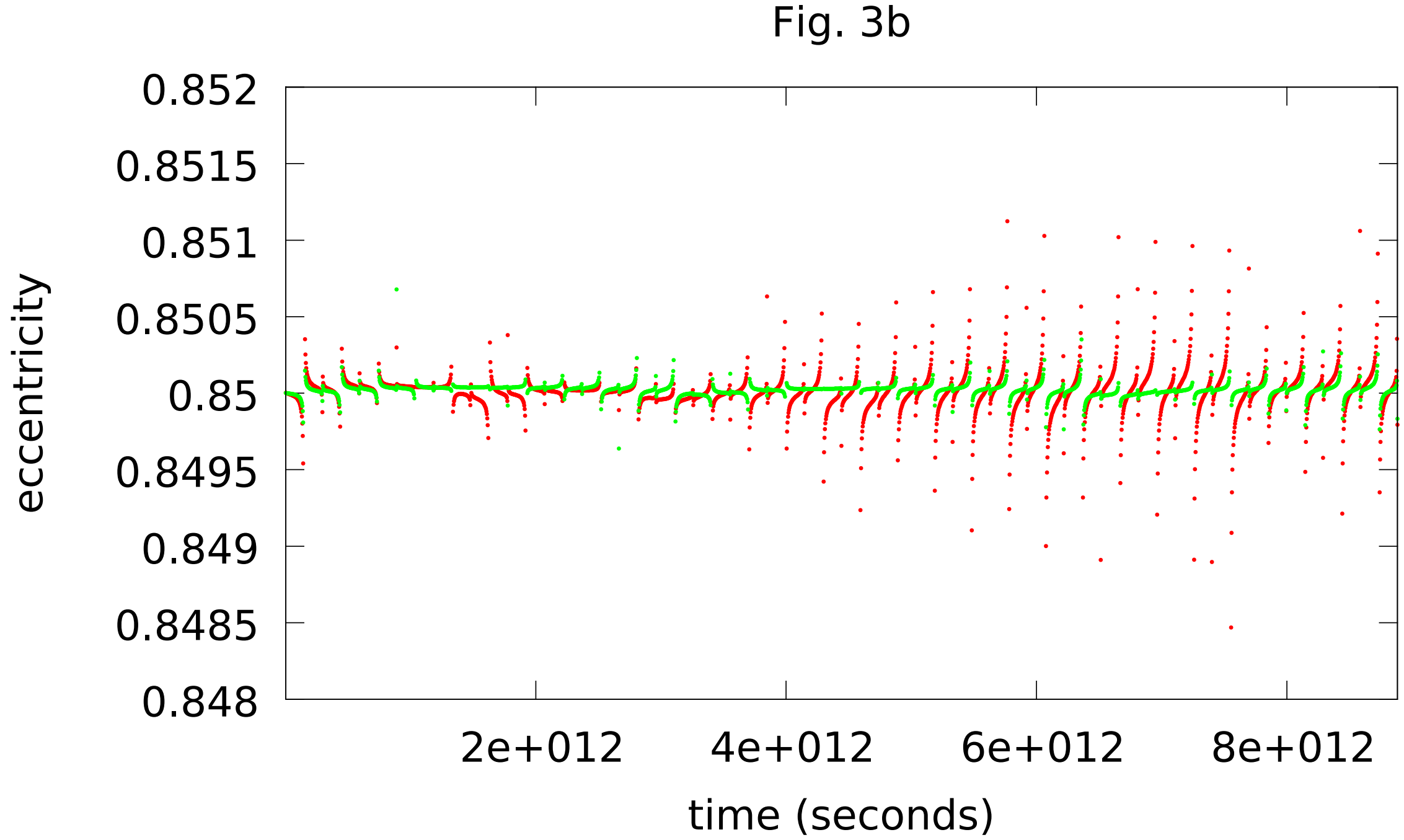

Fig. 3b $ecc = 0.70$ and $\theta = 0 = \phi$ in this figure. Perihelion is $6 \times 10^{13} m$ in red and $1 \times 10^{14} m$ in green.

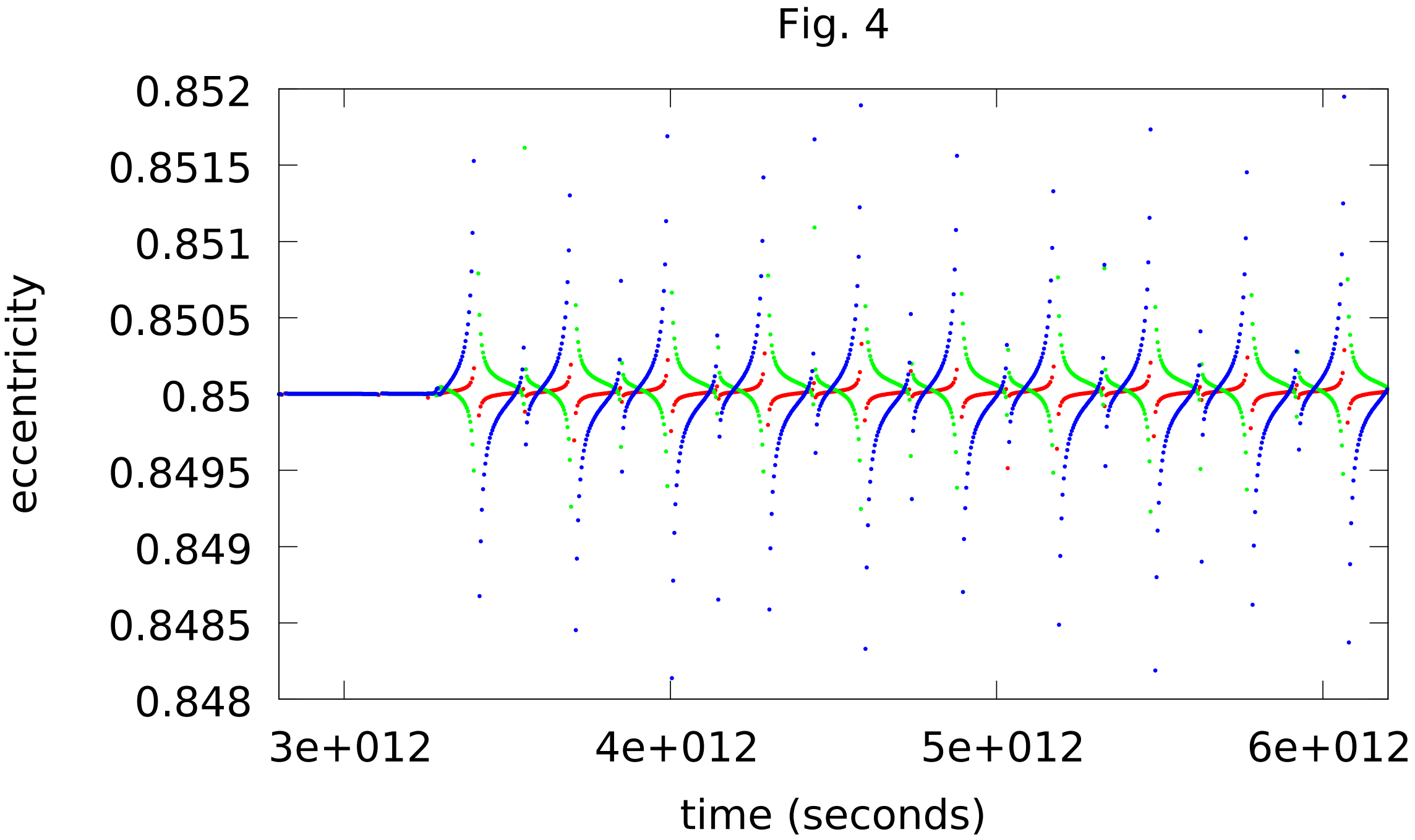

Fig. 4 The WS moves perpendicular to SO's ellipse at a speed of 100 kms/sec and reaches the x-axis when SO has traveled 11.1 cycles. Its initial position perpendicular to its motion is $0.1 \times 10^{16} m$ in red, and $0.05 \times 10^{16} m$ in green. In blue, WS has a mass of $1.0 \times 10^{30} kg$ and $x_0 = 0.05 \times 10^{16} m$.

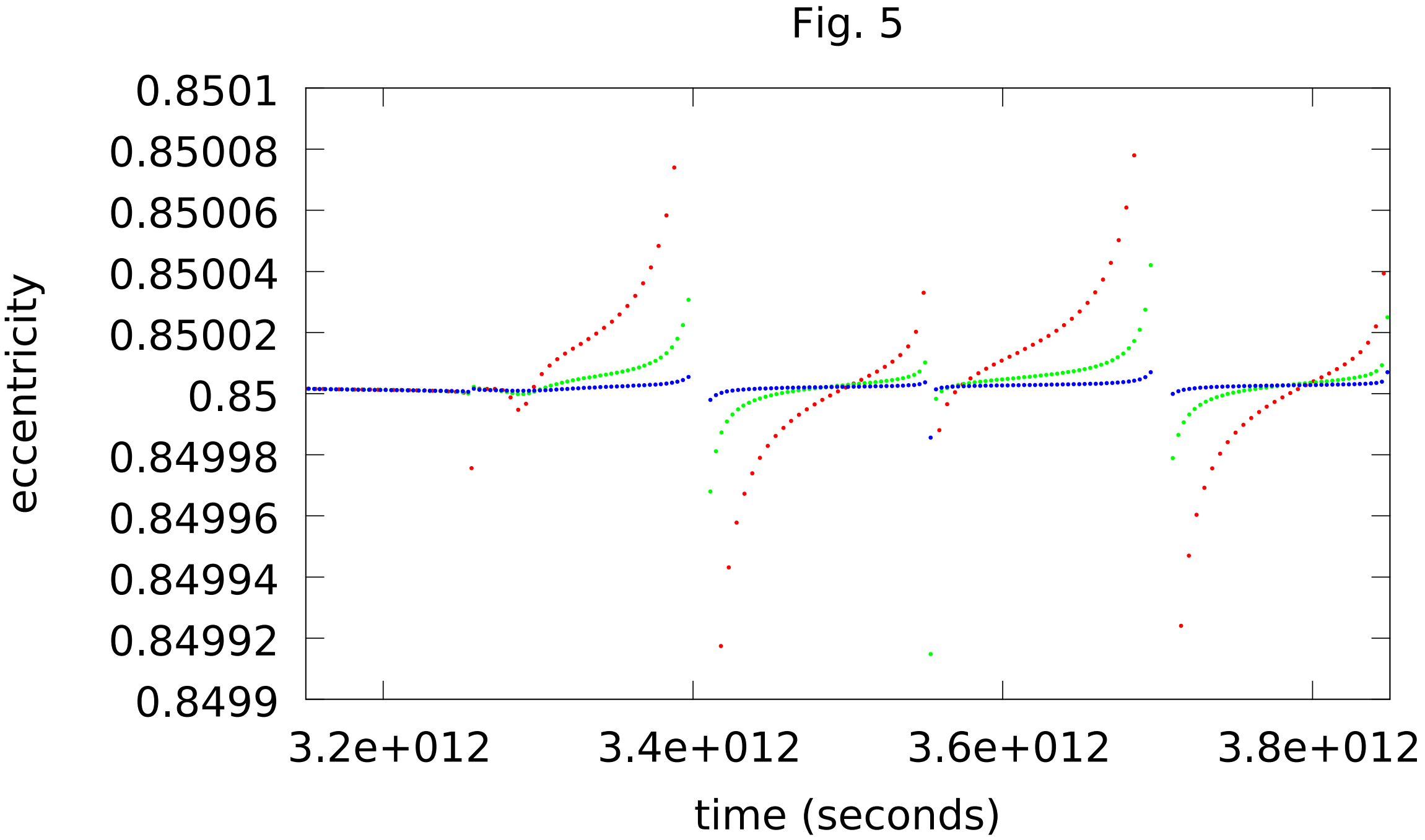

Fig. 5 The WS moves perpendicular to SO's ellipse at a speed of 100 mks/sec, and reaches the x-axis when SO has traveled 11.1 cycles. It has a mass of $0.6 \times 10^{30}$ kg and its initial position perpendicular to its motion is $(0.1, 0.2, 0.4) \times 10^{16} m$ in (red,green,blue)

Fig. 6

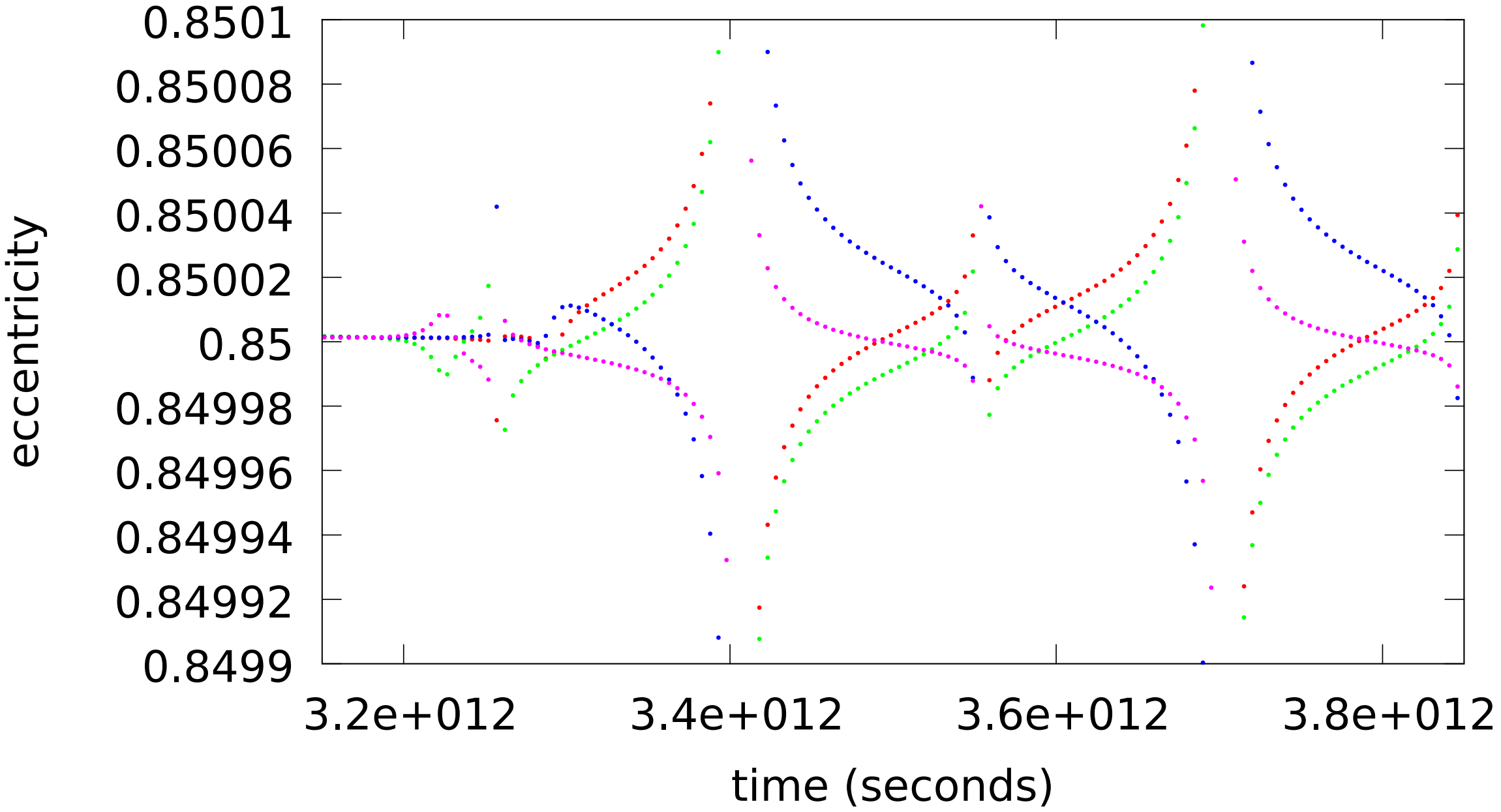

Fig.6 The WS has a mass of $0.6 \times 10^{30} kg$. In red and green, it moves perpendicular to SO's ellipse and starts at $x_0 = 0.1 \times 10^{16}$m and reaches the x-axis when SO has completed 11.1 cycles (red) or 10.9 cycles (green). In blue and magenta, WS moves parallel to SO's ellipse and starts at $y_0 = 0.1 \times 10^{16} m$ and reaches y-axis when SO has completed 11.1 cycles (blue) or 10.9 cycles (magenta)

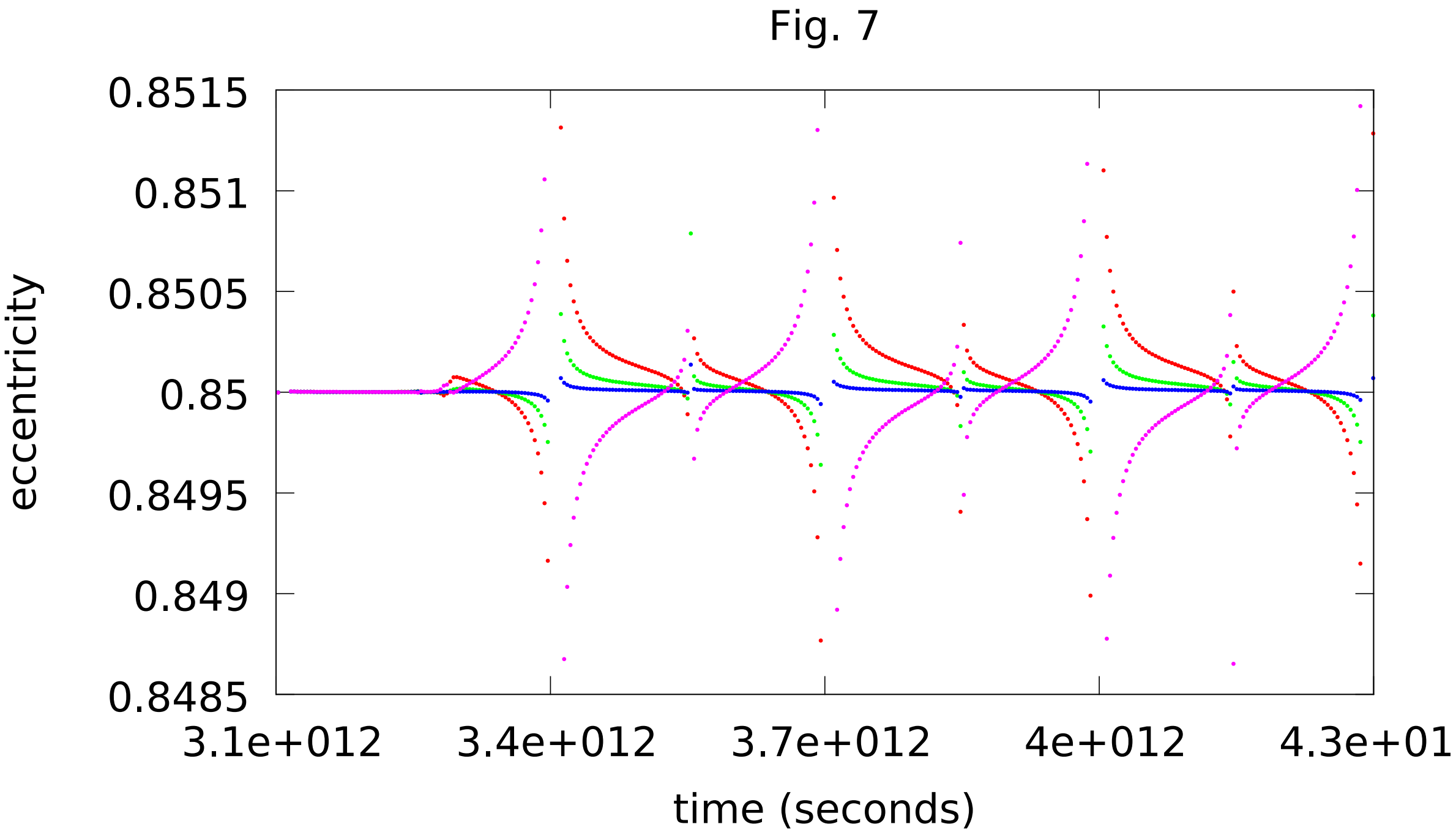

Fig. 7 In this figure, MS has a mass of $1 \times 10^{30} kg$. In red, green and blue it travels parallel, reaches SO when SO has traveled 11.1 cycles and starts at a transverse distance of $(0.05, 0.1, 0.25) \times 10^{16} m$ in (red, green, blue). In magenta, WS travels perpendicular to the ellipse, starts at an initial transverse distance of $0.05 \times 10^{16} m$ and reaches the ellipse when SO has traveled 11.1 cycles.

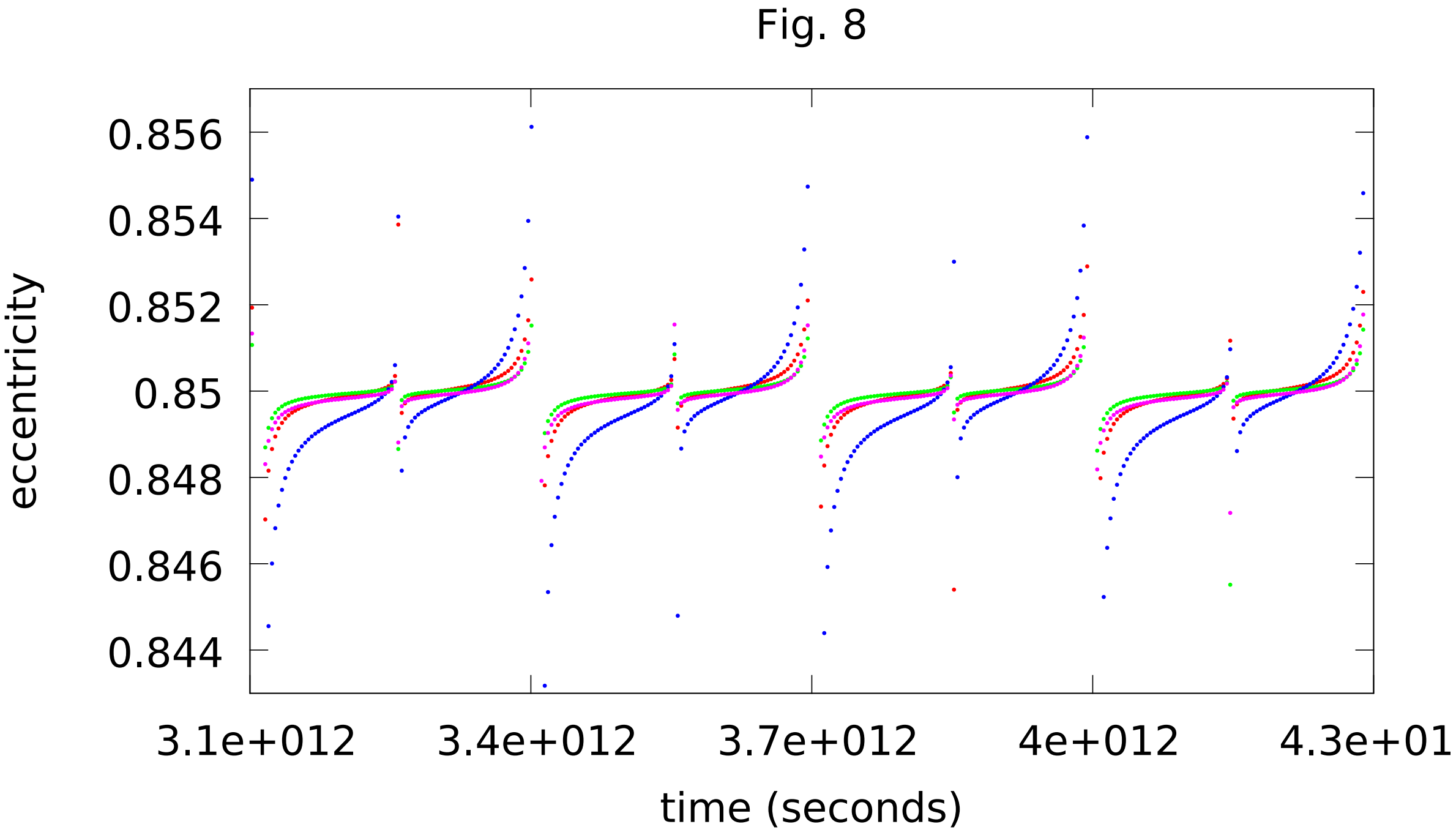

Fig. 8 In red, $(0.6 \times 10^{30}, 0.05LY, 50km/sec)$ in green $(0.3 \times 10^{30}, 0.05LY, 50km/sec)$, in blue $(0.6 \times 10^{30}, 0.05LY, 25km/sec)$, in magenta $(0.6 \times 10^{30}, 1.0LY, 25km/sec)$ with $(mass, y_0, v)$